# DLACB: Deep Learning Based Access Control Using Blockchain

Asma Jodeiri Akbarfam [1], Sina Barazandeh[2], Hoda Maleki[1], Deepti Gupta[3]
[1]Augusta University, GA, USA [2]Bilkent University, Ankara, Turkey [3]University of Texas at San Antonio, TX, USA
[1]ajodeiriakbarfam, hmaleki@augusta.edu,[2] sina.barazandeh@bilkent.edu.tr, [3]deepti.mrt@gmail.com

*Abstract*—In general, deep learning models use to make informed decisions immensely. Developed models are mainly based on centralized servers, which face several issues, including transparency, traceability, reliability, security, and privacy. In this research, we identify a research gap in a distributed nature-based access control that can solve those issues. The innovative technology blockchain could fill this gap and provides a robust solution. Blockchain's immutable and distributed nature designs a useful framework in various domains such as medicine, finance, and government, which can also provide access control as opposed to centralized methods that rely on trusted third parties to access the resources. In existing frameworks, a traditional access control approach is developed using blockchain, which depends on predefined policies and permissions that are not reliable. In this research, we propose DLACB: Deep Learning Based Access Control Using Blockchain, which utilizes a deep learning access control mechanism to determine a user's permissions on a given resource. This proposed framework authenticates the users and logs the access requests on the blockchain to recognize malicious users. The results show that this proposed framework operates correctly for all possible scenarios.

## I. INTRODUCTION

Deep learning has been used in most sectors, including healthcare, autonomous vehicles, farming, finance, etc., to provide better outcomes. Developing a deep learning-based model requires a tremendous amount of data, usually captured by user behavior, Internet of Things (IoT) devices, and sensors. In most scenarios, data is stored at a centralized server, and the model is trained locally to predict the outcome. The centralized deep learning models face a single point of failure and data alteration issues, and data integration issue can corrupt the deep learning training model. It also faces security, data privacy, geo-location, and device heterogeneity challenges, so a centralized model is not considered an ideal solution. To fill the research gap, a decentralized deep learning model [1] was introduced, which showed promising results.

Recent research [2]–[4] presents deep learning-based access control models for various domains; however, it lacks proper authentication of participants. Developing the deep learning-based access control model requires many participants or data sources; an adversary can become a participant or source of data to inject malicious information to corrupt the model. In this research, we introduce the blockchain concept using a deep learning-based access control model to protect the model from the adversary.

Blockchain is a peer-to-peer network [5] and a method of storing information. It prevents unwanted modification and manipulation of data while providing a decentralized ledger for transactions. It is also used for authenticating, maintaining, and synchronizing the content of a transaction ledger replicated across multiple users. Blockchain facilitates decentralized transactions and data management and offers an environment where no third party controls the data, and no trust is required between stakeholders [6]. The features of blockchain make it tamper-resistant, providing integrity and a transparent view of what is happening to the data [7] [8]. Therefore, blockchain is useful in various domains for ensuring security and privacy [6] such as medical systems [9], internet of things [10], finance [11] and access control [12].

Access control restrains access to the system's resources, which is a fundamental security component. Numerous access control methods have been used in blockchain, such as attribute-based access control (ABAC) [13] or role-based access control (RBAC) [14], which require the organization to engineer some roles and attributes in order to have the access control procedure function correctly. Defining and storing different permissions and policies can lead to issues such as the need to add more access, not being capable of making new decisions, or utilizing a large amount of memory in the system [15]. In addition, methods that require third parties to control access suffer from serious drawbacks such as privacy leakage and poor performance in multi-administrative domains and other domains, including users of applications accessing services by sharing their identities. The properties of blockchain-based access control mechanisms have made it possible to address most of the mentioned concerns in various domains [16].

We propose Deep Learning Based Access Control Using Blockchain (DLACB), a framework that provides features according to the organizational needs for accessing data electronically and managing resource access correctly. DLACB is a framework that authenticates users, solves the authentication issue, and is secure against unauthorized access. It utilizes a deep learning model for determining the level of access for each user in a system. This framework is based on a private blockchain in which only trusted nodes can validate the transactions, and a request for resource access is made through the generation of new transactions by the users. DLACB logs the users' requests on the blockchain, which leads to access transparency and helps detect malicious user activities in retrieving resources.

The main contributions of this paper are as follows.

- We identify the research gap in traditional access control models.
- We propose a Deep Learning Based Access Control using Blockchain (DLACB), which solves the authentication issue of multiple participants and does not allow the participants to train the model.
- We present the DLACB process to develop this framework and its cryptographic techniques with details.
- We also present the security analysis and implementation of our proposed work as proof of concept.

The rest of this paper is structured as follows. Section II discusses an overview of blockchain and access control. Section III explores the related work. Section IV discusses the importance of the method. Section V describes the framework and the communication between entities. section VI describes the details of the protocols, Section VII analysis the security of the platform and some design decision, section VIII illustrates how the system was implemented. Section IX explains the limitations of the proposed approach and the paper concludes with section X.

## II. BACKGROUND

### A. Blockchain

Blockchains are peer-to-peer networks and times-tamped chains of blocks maintained by participating nodes and transactions that are accumulated in blocks. Every block is chained to the previous block and cryptographically linked by including the previous block's hash value [17]. Blockchain provides immutable data storage as new blocks which can only be appended to the end of the chain, meaning existing transactions cannot be updated or deleted. Thus, transactions can be trusted without the assistance of any third parties. The immutable chain of historical transactions guarantees non-repudiation of historical transactions. In order to prove identity and authenticity and secure read and write access to the blockchain, cryptography primitives such as digital signatures, hash functions, and ciphers are used. A vital aspect of a blockchain network is the consensus algorithm that ensures integrity and security. The consensus is a procedure that enables all participants in a distributed computing environment to reach an agreement on the ledger's current data state and be able to trust peers [18]. Furthermore, fraud and data tampering is prevented because data cannot be changed without the permission of a majority of the stakeholders in a blockchain system. Blockchain ledgers allow for sharing but not for modification.

There are various types of blockchains, including public and private [19] [20] [21]. A distributed ledger system without constraints and permissions is known as a public blockchain, with examples of Bitcoin and Ethereum. Anyone with internet access can join the network as an authorized node and become a part of the blockchain. It is permitted for a node or user who is a part of the public blockchain to view recent and old records, confirm transactions and engage in mining. The mining and trading of cryptocurrencies is the most fundamental usage of public blockchains.

A restricted or private blockchain [21] [22] that may only be used in a closed network is a private blockchain. A private blockchain is typically utilized within businesses or organizations where only a small group of people are allowed to participate in the blockchain network. The governing organization controls the level of security, authorizations, permissions, and accessibility. In other words, private blockchains are similar to public blockchains, but with restricted access [23]. Private blockchains are mostly used for cases where centralized control of the network is needed, such as asset ownership, digital identity, supply chain management, voting, etc.
Some blockchains, including Ethereum-based ones, consist of smart contracts, which are modular, reusable, and automatically executed programs [24]. A smart contract responds to information sent to it, receives and stores the values, and transfers the information and values to other participants in the blockchain network.

Once a smart contract is launched, the compiler generates a byte code that is stored on the blockchain, and it is assigned a storage address. All network nodes execute the opcodes produced by compiling the contract scripts when a condition specified in the contract takes place. The execution results are then written to the blockchain by the network nodes once a transaction is sent to the contract address.

*B. Access control*

The concept of access control refers to a way of controlling who can view and use resources by defining restrictions and permissions on a system. It is an essential concept in security that minimizes risks such as data breaches to organizations or businesses. The main goal of access control is to decrease the threat of unauthorized access, and security compliance programs rely on access control to protect confidential information, such as customer data, through security technology and access control policies. There are a couple of traditional access control types, including attribute-based access control (ABAC) [25], role-based access control (RBAC) [26], and rule-based access control [27]. In attribute-based access control, every user has an attribute that the administrator defines, and their request can be denied or accepted based on these attributes [28] [29]. In role-based access control, the level of access is defined by the responsibilities and departmental jobs. In rule-based access control, different parts of a system can be accessed based on the predefined rule for that area.

III. RELATED WORK

In this section, we briefly summarize the relevant research on blockchain related access control systems. Blockchain is used for sharing images such as [30], which is proposed for image-based plant phenotyping, and [31], which is for sharing images in medical fields. In these papers, the data owners and patients are in charge of their data and managing their permissions.

In [32], a decentralized data management model is proposed, allowing the patients to control their data. The patient is responsible for remaking a re-encryption key for them. Alhajri et al. [33] propose a consent mechanism based on blockchains for fitness data sharing. The proposed framework is a human-centric decentralized consent system based on blockchains. A medical data sharing and privacy-preserving eHealth system (SPChain) is proposed in [8]. They also use proxy re-encryption schemes allowing patients to share their data while having privacy.

In [34] and [35], blockchain is utilized for access management for IoT, which employs attribute-based access control. [34] considers attribute authorities and IoT devices as two main entities and stores the distribution of attributes in the blockchain, while [35] utilizes smart-contract-based access control for data sharing.

In [36], role-based access control approach control on a blockchain is utilized to share filmed images in closed circuit television installed on roads for crime prevention. [14] uses role-based access control and defines a smart contract-based authentication method appropriate for the trans-organizational utilization of roles. In addition, several security models are discussed in [37]–[45].

In the previous methods, the patient is in charge of granting access to their data which leads to problems such as key management, lack of deployability, authentication rules, and granting incorrect access upon a request. In other platforms, such as [14], traditional access control is used to grant data access, which can lead to many errors in policies or require a considerable amount of memory for storing the access lists. These methods are incapable of dealing with requests which are not seen earlier in the organization, leading to wrongfully denying or accepting an access request.

The DLACB framework authenticates the users and grants correct permissions using deep learning without needing a pre-defined set of policies in the memory. It also logs the access patterns, which leads to recognizing malicious attempts. However, the deep learning model requires a training set that includes all the previously logged access requests and the corresponding response to each request. Using the training set, the model is able to predict the decision for a user's request for a new resource. This is done by the capacity of the deep learning model to capture the hidden similarities of the resources and users based on the previous access logs.

IV. PRIVACY-PRESERVING SYSTEM

In organizations, outsourcing access control functions to third parties has become popular. This eliminates the need to configure and maintain complex systems that could incur high acquisition and operating costs [46]. These systems require log production that usually reveals the system's regular events and assists in detecting malicious behavior or attacks. Analyzing these logs helps to catch security breaches or incident discovery and subsequent damage control [47]. Therefore, these logs need to be immutable to prevent attackers from altering them and causing analysis to be incorrect.

This paper proposes a deep learning access control blockchain framework; a novel approach for implementing access control services by utilizing blockchain technology. Our framework creates immutable system logs and eliminates the need for trusting a third party while carrying the mentioned advantages.

As part of our approach, an access control service is built on top of a private blockchain. It leverages the blockchain technology for configuring access control policies based on a decision engine. In other words, the decision engine evaluates an access request for a resource.

This decision engine is a combination of a deep learning model and priority rules. The use of deep learning for determining access eliminates the need to configure pre-defined policies and rules. It also increases the accuracy of access decisions that have not emerged before. The priority rules are related to conditions in a specific network or organization that must be executed, causing the access control policy to adapt to particular pre-specified circumstances.

## V. DLACB Framework

At a high level, the DCLAB framework uses blockchain to manage access control policies and data retrieval; it first authenticates the user and then provides the user with the appropriate level of access to the requested resources. It also keeps logs of the requests and their results, which are immutable based on the nature of the blockchain.

As illustrated in Figure 1, our framework consists of four entities; user, node, decision engine, and storage. *User* is an entity, such as an employee, interested in storing, manipulating, or accessing data and interacts with our application. A *node* is an entity that runs the blockchain protocol, and *Storage* is an entity that stores the user's data. The storage interacts with the blockchain when a request for storing or retrieving data is made by a user. This interaction is used to correlate data used by the blockchain for consensus purposes. Finally, the *Decision engine* is the core of the DCLAB access control. It authenticates the users and determines the access level of a specific user through a deep learning model.

We use a private blockchain to provide access control and secure the transfer of data between entities. The blockchain contains two types of nodes: user nodes and validator nodes. *User nodes* can generate transactions to be added to the network and are not viewed as trusted entities. The user node can be an automated participant or a portal back-end that is used by a human. A *validator node*, on the other hand, is a trusted entity that receives transactions from user nodes, validates them, and generates blocks to be added to the blockchain. The network requires each validator node to verify its identity before it can participate in the framework as a trusted node. Within our framework, validator nodes are considered standalone servers that are physically secured.

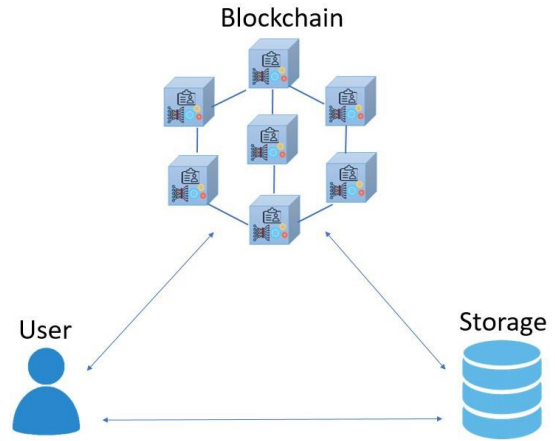

Fig. 1: DLACB framework

For the consensus algorithm, we use Proof-of-Authority (PoA), allowing only the validator nodes to validate transactions and create blocks in the blockchain. Our framework design is as follows. As shown in table I, five types of transactions can be performed on the blockchain: $T_{Setup}$, to add a new node to the system; $T_{AccReq}$, to request access to a resource;

$T_{Link}$ to provide a link to the user; $T_{Storage}$ to log user access to the resource.

$T_{Verified}$ which is created as an input to the smart contract on the decision engine.

Every node on the blockchain has a public and private key pair to encrypt and sign the transactions. For acquiring access to the resources, upon a user request, the blockchain authenticates the user and determines the level of access by utilizing a decision engine. It sends the result to the storage for obtaining an access link which is eventually shared with the user.

## A. Processing data retrieval requests

As illustrated in Figure 2, at a high level, when a user requires access to a resource, it sends the $T_{AccReq}$ to the blockchain, which contains the user's public key, information about the request, and freshness. The blockchain authenticates the user and utilizes the decision engine to determine the user's permissions. The result is sent to the storage. Based on the result the storage responds with $T_{Link}$ which is an access link, timestamp, and nonce that is encrypted with the user's public key. This information is then shared with the blockchain which forwards it to the user.

The user can access the resource on the storage when it receives the link and the nonce.

In the end, the storage sends the $T_{Storage}$ transaction for the blockchain to log the access.

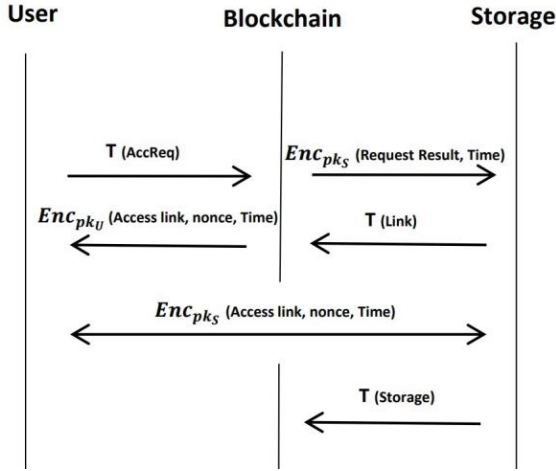

Fig. 2: Requesting data from storage by user

## B. Decision engine

For authentication purposes and to decide the results of requests, the blockchain contains a decision engine. The decision engine is built of two smart contracts, which are stored on the blockchain and are executed when the appropriate transactions are created. Once a data retrieval request is received ($T_{AccReq}$), the *authentication smart contract* is triggered. This smart contract verifies the user's public key and ensures that the user is part of the network. In the case of successful authentication, it creates $T_{Verified}$, which triggers the *authorization smart contract*. The authorization smart contract determines whether the user can access a resource. More specifically, the exciting deep learning model inside the contract will produce an output that contains a set of operations available on the specified resource based on the provided inputs. In the end, a comparison is then made between the list and the priority rules within the smart contract storage, determining if the organization has a specific preference for an event.

TABLE I: Transactions

| Transaction | Parameters |
|---|---|
| $T_{Setup}$ | $pk_A, pk_U, Time, Sign_{sk_A}$ |
| $T_{AccReq}$ | $(pk_U, Time, ReqInfo), Sign_U$ |
| $T_{Link}$ | $Enc_{pk_U}(Accesslink, Nonce, Time), Sign_S$ |
| $T_{Storage}$ | $(Nonce, Time, pk_U), Sign_S$ |
| $T_{Verified}$ | $(Time, pk_{U_{binary}}, ReqInfo_{binary})$ |

## VI. DLACB PROCESS AND CRYPTOGRAPHIC TECHNIQUES

In this section, we will introduce cryptographic notions and techniques, underlying protocols, the details of the transactions, and smart contracts.

### A. Cryptographic notions

We will use standard cryptographic building blocks, including asymmetric encryption, digital signatures, and a cryptographic hash function. The encryption scheme consists of a tuple (*Gen, Een, Dec*)– a generator, an encryption algorithm, and a decryption algorithm respectively. The digital signature scheme is defined by the tuple (*Gens, Sig, V er*)– a generator, a signature, and a verification algorithm respectively. We denote the cryptographic hash function as H and call it a hash function. Finally, we represent the blockchain memory as M; a key-value structured database that contains the complete information stored on the blockchain.

### B. Protocols of DLACB

Our approach contains an auxiliary function, *Access Verification*, and two smart contracts, *Authentication* and *Authorization*.

Algorithm 1 extracts the keys, timestamps, and signatures from the transactions. By checking the signatures and timestamps, the function establishes if the node or the user is the claimed entity and if the transaction is fresh.

Algorithm 2 is the first smart contract performed in the decision engine, the smart contract for authentication is executed because of $T_{AccReq}$. It extracts the user key from the transaction and uses *AccessVerification()* to verify the user and to determine if they were previously added and had the authority to access the system. After the user is verified and the freshness of the request is evaluated, the smart contract converts the user data and resources into a binary representation and issues $T_{Verified}$, which triggers the authorization smart contract. After receiving $T_{Verified}$, algorithm 3 is executed on the decision engine to determine access according to the deep learning model for the user, and then compares the output with the priority rules and sends the final decision to the storage.

**Algorithm 1** Access Verification Check
```
procedure ACCESSVERIFICATION(T_AccReq, M)
    a ← 0
    (pk_U, sign, time) ← Decompose(T)
    if (
        H(pk_c) ∈ M[key] AND
        Current(time) AND verify(time)
    ) {
        a ← 1
    }
    return a
end procedure
```

**Algorithm 2** Authentication
```
procedure VERIFICATION(T_AccReq, M)
    (pk_U, time, ReqInfo, sign) ← Decompose(T_AccReq)
    if (AccessVerificationCheck(T_AccReq, M)
    ) {
        pk_Ubinary = binaryRepr(pk_U)
        ReqInfo_binary = binaryRepr(ReqInfo)
        T_Verified ← (time, pk_Ubinary, ReqInfo_binary)
        return T_Verified
    }
end procedure
```

**Algorithm 3** Authorization
```
procedure DECISION ENGINE(T_Verified, M)
    (time, pk_U, ReqInfo) ← Decompose(T_Verified)
    if (Current(time)
    ) {
        accessList ← DecisionEngine(pk_U, ReqInfo)
    }
    (storage) return accessList
end procedure
```

## VII. SECURITY ANALYSIS

In this part, we analyze the features of DLACB, which contribute to securely sharing data with the users.

*Confidentiality and integrity*: Each entity on the blockchain has a pair of public and private keys, and requests exchanged between two framework entities are encrypted with the receiver's public key. For example, the $T_{Link}$ sent by the storage is encrypted with the user's public key, whereas the Request Result is encrypted with the storage's public key; This ensures confidentiality and prevents an attacker from comprehending the content of a transaction or a request. DLACB also ensures integrity by demanding the entities which are sending a request to sign the content with their private key. The signature proves that the request originated from the entity and was not modified.

*Preventing reply attack*: For each request and produced link, a nonce which is a randomly-generate token is issued. The nonce ensures that a malicious user can not reuse previously communicated links, consequently preventing reply attacks.

*Preventing man in the middle attack*: There are different scenarios in that, a man-in-the-middle attack is possible but prevented by encryption of the data, including the nonce and the signature of the message inside the request or the transaction.

*Detecting malicious activity*: The framework keeps records of the user's requests to access specific data as well as their permission to access them. This process is done by logging the requests and the results on the blockchain, which can detect malicious requests and detect attacks. In addition to the mentioned features, the DLACB uses PoA as a consensus algorithm, in contrast to Proof-of-Work consensus, where an attacker needs to control 51% of the network's computing power; in PoA consensus, it is necessary to control 51% of the network nodes. It is more difficult to control the nodes in a permissioned network than it is to acquire computational power, leading the attack to be more challenging in PoA [48].

PoA is scalable, efficient, and easy to implement; it can be scaled so that a blockchain using our model can have $v$ validators with $v > 2$. Validators, the trusted nodes, are the only nodes responsible for reaching consensus, while the users are free to submit transactions.

## VIII. IMPLEMENTATION

This section discusses the implementation details of the introduced DLACB framework, including the blockchain and the Deep-Learning-based decision engine, and how these components work together as a system. In the end, we provide details on the experimental results and discuss the practicality of the framework from a technical perspective.

### A. Blockchain Implementation

DLACB is proposed primarily for organizations that use private networks which require particular transactions, authorization procedures, and smart contracts, as discussed in section V. Therefore, the blockchain is implemented from scratch, not inherited from any off-the-shelf blockchain technology, and can be considered a baseline for future frameworks using Deep Learning for the Access Control decision engine. To implement the framework as a baseline, as mentioned and as proof of concept, we use Python3 as the backend language. The framework's functionality is implemented as an API using the Flask library. In addition, a simple interface is implemented using plain HTML styled with CSS, and JavaScript is used to connect the interface and the backend by sending and receiving the HTTP requests. Different ports are used for trusted nodes and users, and the blockchain runs on a defined port necessary for communication through the API.

The user's public and private keys are generated as pairs, and their uniqueness is ensured. Here we use the widely used SHA-256 cryptographic hash function [49] as our encryption method for the entity and verification of digital signatures. The transactions are implemented as mentioned in section V, and the transactions are then extracted and distributed among the nodes for consensus. The keys and the data are stored locally by serializing and restored by de-serializing. This method is only suitable for the demo version. However, the deployed version of the framework requires the use of a DBMS database management system such as MySQL.

### B. Decision Engine Implementation

In this work, we use a machine learning model as our decision engine to allow or reject access to a resource upon a user's request. For this purpose, we use DLABCAplha [15], a deep learning model inspired by the widely used [50] model, a convolutional Deep Neural Network successfully used in various domains. The model is implemented using popular Tensorflow and Keras machine learning libraries in python 3.9. We use a pre-trained version of the model publicly available on GitHub. The model receives the user identification, which is an *id*, resource *id*, and returns four values for all the possible operations the user might have requested for the specific resource. The decision is made based on the requested operation. Nevertheless, the model is not limited to the use of *user$_{id}$* and *resource$_{id}$* and can function if trained on any attribute of the users or resources as well as user roles. The size of the trained decision engine and the weights of the model is a small 1 MB file and has little computational and memory overhead.

Deep learning models can be trained by utilizing any numerically representable feature. Therefore, Depending on the use case, the attributes utilized to train the model can be derived from attributes stored in a medical facility or, in general, an access control system containing users or resources with various attributes.

Nevertheless, a deep learning model does not consider the predefined rules of the blockchain for accessing resources. To address this, we pass the request through a rule-checker, a set of static rules defined to ensure the validity of the decision made by the engine. The rule-checker consists of rules that apply to the user-resource pairs in specific cases. Although the deep learning model is capable of generalizing decisions, it is not suitable for particular cases and static rules, and the existence of the additional layer is necessary.

### C. Experimental Results

To prove the practicality of the developed framework, we run experiments to ensure it performs as expected in all possible scenarios. The results of the experiments and the requests, in general, are logged in the blockchain. The scenarios include the following cases:

1) A user attempts to access a resource while the user and the public key are not registered on the blockchain. In this scenario, the access control method should deny the request.
2) The user accessing the resource is authenticated in the system, but the deep learning model predicts that the user's request should be denied.
3) After the deep learning model has determined the user's request as accepted, the policies on the decision engine will deny the request. This denial is based on static rules and policies defined by the organization's administrators.
4) Last but not least is where the decision engine and the policies agree on authorizing the user to access the resource.

We test the model with all of the above scenarios, and the framework completes every test successfully. For testing purposes, only 100 users and the corresponding transactions are registered on the blockchain, while the engine recognizes the entire dataset, including all users. The test cases for the scenarios are selected carefully to satisfy the description of each case. The lightweight model has been tested both locally and remotely, assigning a specific range of ports for users and admins.

## IX. LIMITATIONS

The framework provides several benefits over the state-of-the-art, but there are also some limitations to be considered. A machine-learning model is used to train the decision engine, which uses the logged access requests and corresponding responses for training. Consequently, it is necessary to train the model for samples including all users so that they can be correctly identified with their IDs or attributes. Among the challenges presented are that the model checkpoint must be stored and updated later for new users, which requires additional computations. Although the framework requirement is not desirable, the computation required to update the model will not take long and is insignificant. In addition, the model requires the ability to identify each user's access to different resources, either by adding a user to the model using their attributes, which will assist the decision engine or by identifying the resources that are accessible or restricted to each user. It is important to note that DLACB is designed for private purposes, and such logs or attributes are expected to exist, and it focuses on a general access control approach. Hence, the blockchain and the decision engine can be extended to specific domains, such as e-health or e-government, and modified based on their needs.

## X. CONCLUSION

Maintaining logs of access control patterns for detecting malicious behavior and determining the proper amount of access for the users is essential. In access control methods such as ABAC and RBAC, the administrator engineers attributes, roles, and rules for applying access control. The administrator also ensures the policies are suitable for each organization and each user has the correct amount of access to the resources, which can lead to many errors or insufficient or extensive access of some users to the resource. These methods work for predefined cases within a broader organization; therefore, they can not make the correct choice when encountering a new issue [51]. So consequently, there is a necessity for an alternative access control method.

In this paper, we presented DLACB, a blockchain-based access control platform that authenticates the users and utilizes a decision engine that employs a deep learning model along with priority rules to determine the extent of access required upon a user request. We implemented DLACB on a private blockchain and utilized this framework to evaluate all the possible access scenarios, indicating that the platform successfully authenticates the users, logs the access patterns, and determines the correct level of access upon access requests.

## XI. ACKNOWLEDGEMENT

This work was funded by NSF grant CCF-2131509 and Augusta University Provost's office and the Translational Research Program of the Department of Medicine.